\documentclass{ws-mpla}
\usepackage[super]{cite}
\usepackage{graphicx}
\begin{document}

\def\met{\displaystyle{\not}E_T}
\let\jnfont=\rm
\def\NPB#1,{{\jnfont Nucl.\ Phys.\ B }{\bf #1},}
\def\PLB#1,{{\jnfont Phys.\ Lett.\ B }{\bf #1},}
\def\EPJC#1,{{\jnfont Eur.\ Phys.\ Jour.\ C }{\bf #1},}
\def\PRD#1,{{\jnfont Phys.\ Rev.\ D }{\bf #1},}
\def\PRL#1,{{\jnfont Phys.\ Rev.\ Lett.\ }{\bf #1},}
\def\MPLA#1,{{\jnfont Mod.\ Phys.\ Lett.\ A }{\bf #1},}
\def\JPG#1,{{\jnfont J.\ Phys.\ G }{\bf #1},}
\def\CTP#1,{{\jnfont Commun.\ Theor.\ Phys.\ }{\bf #1},}
\newcommand\JHEP[3] {{\it JHEP\ }{\bf #1} (#2) #3}

\markboth{X. Gong, Z. -G. Si, S. Yang and Y. -J. Zheng}
{Determination of charged Higgs couplings at the LHC}

\catchline{}{}{}{}{}

\title{DETERMINATION OF CHARGED HIGGS COUPLINGS \\AT THE LHC}

\author{\footnotesize XUE GONG}

\address{School of Physics, Shandong University\\
Jinan, Shandong 250100,  China\\
gongxue@mail.sdu.edu.cn}

\author{ZONG-GUO SI}

\address{School of Physics, Shandong University\\
Jinan,
Shandong 250100,  China\\
Center for High Energy
Physics, Peking University\\
Beijing 100871, China\\
zgsi@sdu.edu.cn
}

\author{SHUO YANG}

\address{Physics Department, Dalian University\\
Jinan,
Dalian, 116622, China\\
yangshuo@dlu.edu.cn
}

\author{YA-JUAN ZHENG}

\address{CASTS, CTS and Department of Physics, National Taiwan University\\
Taipei 10617, China\\
yjzheng218@gmail.com
}

\maketitle


\begin{abstract}
We review the study of the charged Higgs and top quark associated production at the LHC with the presence of an additional scalar doublet. Top quark spin effects are related to the Higgs fermion couplings through this process. The angular distributions with respect to top quark spin
turns out to be distinctive observables to study the $Htb$ interaction in different models.
\end{abstract}

\ccode{PACS Nos.: 14.80.Cp,14.65.Ha,12.60.-i}

\section{Introduction}	

Gauge symmetry and electroweak spontaneous symmetry breaking (EWSB) are two fundamentals of the Standard Model (SM) of particle physics. The discovery of gauge bosons $W$ and $Z$ and further measurements established the gauge group $SU(3)_{C}\times SU(2)_{L}\times U(1)_{Y}$  of the SM.  While the mechanism of EWSB is implemented by introducing one complex Higgs doublet $\Phi$ and then triggered by the development of neutral component vacuum expectation value (VEV). Weak gauge bosons and fermions get their masses in this way.  In July 2012, a SM like Higgs boson with mass around 126 
GeV has been discovered at the LHC 
by ATLAS and CMS collaborations\cite{Higgs_ATLAS,Higgs_CMS}. However, that introducing only one complex scalar doublet in the SM is simply based on minimal principle. It is thus natural to consider scenarios with additional complex scalars such as the two Higgs doublet model (2HDM).

Depending on Yukawa couplings between Higgs and fermions,  2HDMs are classified as different types\cite{higgshunter'sguide, 2hdm_review}. From experimental aspects, flavour changing neutral current (FCNC) should be highly suppressed at tree-level, which is realized by Glashow-Iliopoulos-Maiani (GIM) mechanism in the SM. However, in 2HDMs with extended Higgs sector, extra discrete symmetry $Z_2$ is usually introduced to suppress tree-level FCNC. After the spontaneous symmetry breaking, there are five physical Higgs scalars,
{\it i.e.}, two neutral CP-even bosons $h_{0}$ and $H_{0}$,
one neutral CP-odd boson $A$, and two charged bosons $H^{\pm}$. In such models with multiple neutral scalars, the mixings between these components will make it difficult to disentangle the Higgs properties. Hence, it is important to study the charged Higgs bosons, which might be able to provide unambiguous signatures to distinguish from models with extended Higgs sector.

Constraints on the charged Higgs boson in the 2HDM are given from both collider and flavour experiments. 
One model-independent direct limit is from the LEP experiments, which gives $M>78.6$ GeV at 95\% C.L. 
through exclusive decay channels of $H^{+}\to c\bar{s}$ and $H^{+}\to\tau^{+}\nu$\cite{LEP_Higgs},
where $M$ represents the charged Higgs mass.
At hadron colliders, the approaches used to search for the charged Higgs are different in the low mass region 
$M<m_t$ and in the large mass range $M>m_t$. 
When $M<m_t-m_b$, charged Higgs can be produced through the top quark decay $t\to H^+b$ followed by decay mode $H^+\to \bar{\tau}\nu$. 
On the other hand, when $M>m_t+m_b$,  the dominant production process through gluon bottom fusion $gb\to tH^-$, followed by dominant decay modes $H^- \to t\bar{b}$ and $H^-\to \tau \bar{\nu}$. 
The Tevatron constrains 2HDM for a charged Higgs boson with the lower bound as $\sim 160$ GeV \cite{Tevatron, D0_0908,
CDF, D0_0906}.  
In addition, the indirect flavour constraints can be
extracted from B-meson decays since the charged Higgs contributes to the FCNC process. 
In Type-II 2HDM, a lower limit on the charged Higgs mass $M>316~{\rm GeV}$ at $95\%$ C.L. 
is obtained mainly from $b\to s\gamma$ branching ratio measurement irrespective of the value of 
$\tan\beta$ \cite{typeII_constraints,2HDM_constraints}.
However, in Type-III or general 2HDM the phases of the Yukawa couplings are free parameters so that
$M$ is less constrained and can be as low as 100 GeV\cite{btosgamma1,ssbao_0801}. For more detailed discussions on phenomenological constraints
on charged Higgs, we refer to Ref. 14.

Along with the experimental search for the charged Higgs boson, there are also extensive phenomenological studies on charged Higgs
boson production at the LHC
\cite{collider_higgs,Deshpande:1983xu, Willenbrock:1986ry, Krause:1997rc, collider_higgs2,Moretti1999,Bao_2012,dominate_process, dominate_process2,Olness:1987ep,tripleb, fourb,pt,Gunion:1993sv,Czarnecki,dominate_process3, huang,Kidonakis:2004ib}.
Especially, the gluon bottom fusion process $gb\rightarrow tH^{-}$
for the charged Higgs mass $M>m_{t}+m_{b}$ 
shows a lot of interesting signatures due to the large couplings of $Htb$ interaction \cite{dominate_process,tripleb,fourb,pt,dominate_process3,Kidonakis:2004ib,H+Jetsub}. 
The next-to-leading order QCD corrections to this process has also been performed in Ref. 34-37. 

In this Letter, we revisit the process $gb\rightarrow tH^{-}$ at the LHC and take a method similar to 
Ref. 38, 39 to optimize this signal from the SM backgrounds. 
As claimed in Ref. 38,
the angular distribution related to top quark spin is efficient to
disentangle the chiral coupling of the $W'$ boson to the SM fermions. 
The left-right asymmetry induced by top
quark spin for the process $pp\to tH^-$ has been analyzed in Ref. 40, 41.
Here we further investigate such a kind of effect after including the charged Higgs and top quark decay, and we employ the angular distributions of the top
 quark and the final state leptons to disentangle $Htb$ couplings at LHC.  

The following discussions are organized as follows. 
In Section \ref{model}., the corresponding
theoretical framework is briefly introduced.
 In Section \ref{seciii},
we perform the numerical analysis of top quark and charged Higgs
associated production process from the gluon bottom fusion. 
Specifically, the correlated angular distributions of the final state particles are
investigated to identify the interaction of top-bottom quark and charged
Higgs. Finally,  summary and discussions are given in Section \ref{summary}.

\section{Theoretical Framework}\label{model}

\subsection{The Lagrangian and constraints on the model parameters }

To begin with, we give a brief introduction to the two-Higgs-Doublet Models (2HDMs), which is  one of the minimal 
extensions of the SM. Different from the SM with only one scalar doublet,  two complex ${SU(2)_{L}}$ doublet scalar fields are introduced in the 2HDMs, which can be written as
\begin{equation}
\Phi_{i}=\left(\begin{array}{c}
H_{i}^{+}\\
(H_{i}^{0}+iA_{i}^{0})/\sqrt{2}\end{array}\right),\end{equation}
 where $i=1, 2$. Imposing CP invariance and ${\rm U(1)_{EM}}$ gauge
symmetry, the minimization of potential gives
\begin{eqnarray}
\langle \Phi_{i}\rangle=\displaystyle{\frac{1}{\sqrt{2}}}\left(\begin{array}{c}
0\\
v_{i}\end{array}\right),\end{eqnarray}
with $v_{i}~(i=1, 2)$ is non-zero vev. 
 After the spontaneous symmetry breaking, there are five physical Higgs scalars,
{\it i.e.}, two neutral CP-even bosons $h_{0}$ and $H_{0}$,
one neutral CP-odd boson $A$, and two charged bosons $H^{\pm}$.
The ratio between the vets of the two scalar doublets is defined as ${\displaystyle {\tan\beta\equiv{v_{2}}/{v_{1}}}}$, which determines the interactions of the Higgs fields with the vector bosons
and fermions. For simplicity, the masses of these Higgs bosons are assumed to be degenerate. 

\begin{figure}
\begin{centering}
\begin{tabular}{c}
\includegraphics[width=0.6\textwidth]{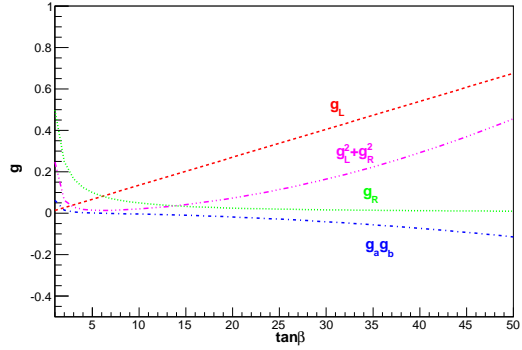} 
\end{tabular}
\caption{Couplings $g_L$, $g_R$, $g_L^2+g_R^2$ and $g_ag_b=\frac{g_R^2-g_L^2}{4}$ versus $\tan\beta$.} 
\label{fig:coupling} 
\par\end{centering}
\centering{} 
\end{figure}

Since phenomenologies from charged Higgs are possible to shed some lights on the extended Higgs sector, in the following we aim to study the charged Higgs signatures at the LHC in the framework of the Type II 2HDM Yukawa couplings,
\begin{eqnarray}
-\mathcal{L} & = & -\cot\beta\frac{m_{u}}{v}\bar{u}_{L}(H+iA)u_{R} +\tan\beta\frac{m_{d}}{v}\bar{d}_{L}(H-iA)d_{R}\nonumber \\
 &  & -\sqrt{2}\cot\beta\frac{m_{u}}{v}V_{ud}^{\dag}\bar{d}_{L}H^{-}u_{R}-\sqrt{2}\tan\beta\frac{m_{d}}{v}V_{ud}\bar{u}_{L}H^{+}d_{R}+{\rm h.c.}.\label{lagrangian}\end{eqnarray}
The VEV of the SM Higgs is related as $v=\sqrt{v_{1}^{2}+v_{2}^{2}}$. 
Following the Ref. 3, $t\bar{b}H^{-}$ coupling can be written as
 \begin{eqnarray}
g_{H^{-}t\bar{b}}
&=&g_{a}+g_{b}\gamma_{5}.
\end{eqnarray}
Within the Type-II 2HDM, 
$$g_{a,b}=g({\rm cot}\beta m_{t}\pm {\rm tan}\beta m_{b})/(2\sqrt{2}m_{{\rm W}}).$$ 
 We can also define $g_L$ and $g_R$
\begin{eqnarray}
-\mathcal{L} & = & g_L \bar{t}_Rb_LH^+ + g_R \bar{t}_Lb_RH^++{\rm h.c.}\\ \nonumber
& = &  \left(g_L \bar{t}\frac{1-\gamma_5}{2}b+g_R\bar{t}\frac{1+\gamma_5}{2}b\right)H^++{\rm h.c.}\\\nonumber
&=&\bar{t}\left(\frac{g_L+g_R}{2}+\frac{g_L-g_R}{2}\right)b H^+ +{\rm h.c.} \\\nonumber
&=&\bar{t}\left(g_a+g_b\right)b H^+ +{\rm h.c.}
\label{lagrangian_1}\end{eqnarray}
Behaviours of coupling constants $g_L$, $g_R$, $g_a$ and $g_b$ are shown versus $\tan\beta$ in Fig.~\ref{fig:coupling}. We can see that the value of $g_ag_b$ becomes minus around $\tan\beta\approx7\approx\sqrt{\frac{m_t}{m_b}}$, while $g_ag_b>0$ corresponds to $\tan\beta<7$, and $g_ag_b<0$ corresponds to $\tan\beta>7$.

\subsection{ $tH^-$ associated production at the LHC}

\begin{figure}[h]
\centering \includegraphics[width=0.7\textwidth]{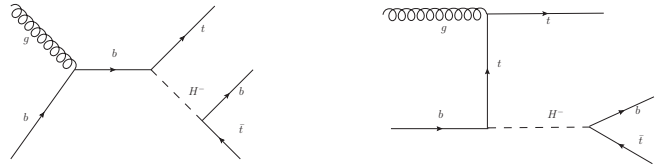}
\caption{Feynman diagrams for $gb\to tH^{-}\to t\bar{t}b$ process. }
\label{fig:com1}
\end{figure}

We begin to consider the following process of charged Higgs production, and the Feynman diagrams are illustrated in Fig.~\ref{fig:com1}.
\begin{eqnarray}
\nonumber
g(p_{1})+b(p_{2})&\to& t(p_{3},s_{t})+H^{-}(p_{4})\\
&\to& t(p_{3},s_{t})+b(p_{5})+\bar{t}(p_{6},s_{\bar{t}}),\label{pp2th2ttb}\end{eqnarray}
 where $p_{i}$ is the four-momentum of the corresponding particle.
$s_{t}$($s_{\bar{t}}$) is the top (antitop) quark spin vector in 4-dimension
with $s_{t}^{2}=s_{\bar{t}}^{2}=-1$ and $p_{3}\cdot s_{t}=p_{6}\cdot s_{\bar{t}}=0$.

When the charged Higgs is produced on-shell , under the narrow width approximation, we have 
\begin{equation}
\lim_{\Gamma\to0}\frac{1}{(p_{4}^{2}-M^{2})^{2}+\Gamma^{2}M^{2}}
\longrightarrow\frac{\pi}{\Gamma M}\,\delta(p_{4}^{2}-M^{2}),\end{equation}
 where $\Gamma$ and $M$ represent the charged Higgs boson decay width and mass, respectively. 
 The squared matrix element for the process (\ref{pp2th2ttb}) including top quark spin information can be written as 
\begin{eqnarray}
\nonumber
|{\cal M}(s_{t},s_{\bar t})|^{2}=&&|{\cal M}_{gb\to tH^{-}}(s_{t})|^{2}
|{\cal M}_{H^{-}\to b\bar{t}}(s_{\bar{t}})|^{2}\times\frac{\pi}{\Gamma M}\delta(p_{4}^{2}-M^{2}),\end{eqnarray}
 where
 \begin{equation}
|{\cal M}_{gb\to tH^{-}}(s_{t})|^{2}=\frac{g_{s}^{2}}{2N_{c}}\Big\{{\cal A}+
{\cal B}_{1}(p_{1}\cdot s_{t})+{\cal B}_{2}(p_{2}\cdot s_{\bar{t}})\Big\},\label{eqst2}\end{equation}
and 
\begin{eqnarray}
|{\cal M}_{H^{-}\to b\bar{t}}(s_{\bar{t}})|^{2}&=&(g_{a}^{2}+g_{b}^{2})(M^{2}-m_{b}^{2}
-m_{t}^{2})-2(g_{a}^{2}-g_{b}^{2})m_{b}m_{t}\nonumber \\
&&-4g_{a}g_{b}m_{t}(p_{5}
\cdot s_{\bar{t}}).\label{eqst1}\end{eqnarray}
 The expressions of ${\cal A}$, ${\cal B}_{1}$ and ${\cal B}_{2}$ are
\begin{equation}
{\cal A}=(g_{a}^{2}+g_{b}^{2})\, A_{1}\,+\, m_{b}m_{t}(g_{b}^{2}-g_{a}^{2})\, A_{2},\end{equation}
 with 
\begin{eqnarray}
\nonumber
A_{1}&=&\displaystyle{\frac{\hat{s}(p_{1}\cdot p_{3})-m_{b}^{2}(4p_{1}\cdot p_{3}+3p_{2}\cdot p_{3})}
{(\hat{s}-m_{b}^{2})^{2}}}+\frac{\hat{s}(p_{1}\cdot p_{3})+m_{t}^{2}(\hat{s}-2p_{2}\cdot
p_{3})}{4(p_{1}\cdot p_{3})^{2}}\nonumber\\
&&-\frac{1}{2(p_{1}\cdot p_{3})(\hat{s}-m_{b}^{2})}\big\{m_{t}^{2}(\hat{s}-2m_{b}^{2})-2(p_{1}\cdot p_{3})m_{b}^{2}\nonumber\\
&&+2(\hat{s}-2p_{2}\cdot
p_{3})(p_{1}\cdot p_{3}+p_{2}\cdot p_{3})\big\},\\
A_{2}&=&\frac{(\hat{s}+2m_{b}^{2})}{(\hat{s}-m_{b}^{2})^{2}}+\frac{m_{t}^{2}-p_{1}\cdot p_{3}}{2(p_{1}
\cdot p_{3})^{2}}-\frac{2p_{1}\cdot p_{3}+4p_{2}\cdot p_{3}-\hat{s}}{2(p_{1}\cdot p_{3})(\hat{s}-m_{b}^{2})},
\end{eqnarray}
and 
\begin{eqnarray}
B_{1}&=&2g_{a}g_{b}m_{t}[\frac{4m_{b}^{2}-\hat{s}}{(\hat{s}-m_{b}^{2})^{2}}+\frac{2p_{2}\cdot p_{3}-\hat{s}}{4(p_{1}
\cdot p_{3})^{2}}+\frac{1}{p_{1}\cdot p_{3}}-\frac{p_{2}\cdot p_{3}}{(p_{1}\cdot p_{3})(\hat{s}-m_{b}^{2})}],\\
B_{2}&=&2g_{a}g_{b}m_{t}[\frac{3m_{b}^{2}}{(\hat{s}-m_{b}^{2})^{2}}+
\frac{m_{t}^{2}-p_{1}\cdot p_{3}}{2(p_{1}\cdot p_{3})^{2}}\frac{\hat{s}-p_{1}\cdot p_{3}-2p_{2}\cdot p_{3}}{(p_{1}\cdot p_{3})(\hat{s}-m_{b}^{2})}].
\end{eqnarray}
 The similar results can be obtained for
the process $g\bar{b}\to\bar{t}H^{+}\to t\bar{t}\bar{b}$ 
from the above equations by using CP-invariance.
We can see that the top quark spin effects are related to 
the product of $g_ag_b$ while disappear for a pure scalar
or pseudo-scalar charged Higgs boson. 
For $pp\to tH^-\to t\bar{t}b$ process, 
 this feature can be reflected by the following spin observable
\begin{equation}
\label{double}
<{\cal {O}}_{t}> =2<\bf {{S_{t}}}\cdot  {\bf \hat a}>= {\sigma(\uparrow) - \sigma(\downarrow)\over
\sigma(\uparrow) + \sigma(\downarrow)},\\
\end{equation}
where $\bf{{S_{t}}}$ is the top quark spin vector in its own rest frame,
and the arrows on the right-hand side refer to the spin state of the top quark corresponding to the quantization unit axis ${\bf \hat a}$.
At the LHC, the helicity basis serves a better choice, i.e.,
${\bf\hat a}={\bf {\hat{p}}}_t^*$ with the 3-momentum unit vector ${\bf {\hat{p}}}_t^*$ in the
$tH^-$ center-of-mass frame. 
We can also define the spin observable of antitop quark in a similar way
\begin{equation}
<{\cal {O}}_{\bar t}> =2<\bf {{S}_{\bar t}}\cdot {\bf \hat b}>,
\end{equation} 
where $\bf{\hat b}$ is the spin quantization axis corresponding to antitop quark. At the LHC, we can choose
${\bf\hat b}={\bf {\hat{p}}}_{\bar {t}}^*$ with the 3-momentum unit vector
${\bf {\hat{p}}}_{\bar{t}}^*$ in the
charged Higgs rest frame. 

For the polarized top quark decay 
$$
t({\bf{S_t}}) \to c(p_c)+X,
$$
where $c$ represents a final state particle or jet
and $p_c$ represents its four-momentum,
the corresponding differential distribution can be written as\cite{Brandenburg:2002xr}
\begin{eqnarray}\label{power}
\frac{1}{\Gamma_t}\frac{d\Gamma_t}{d\cos\vartheta}=\frac{1}{2}\left(1+\kappa_c\cos\vartheta\right),
\end{eqnarray}
where $\vartheta$ is the angle between the top quark spin vector and the direction of 
$c$ in the top quark rest frame. $\kappa_c$ is the spin analysing power of the corresponding particle or jet $c$.
For the charged lepton from the semileptonic top quark decay within SM,  
we have $\kappa_{l^+}=1$ at the tree level.

\section{Numerical Results and discussion}
\label{seciii}
%
\begin{figure}[h]
\begin{centering}\rotatebox{270}{
\begin{tabular}{c}
\includegraphics[width=0.30\textwidth]{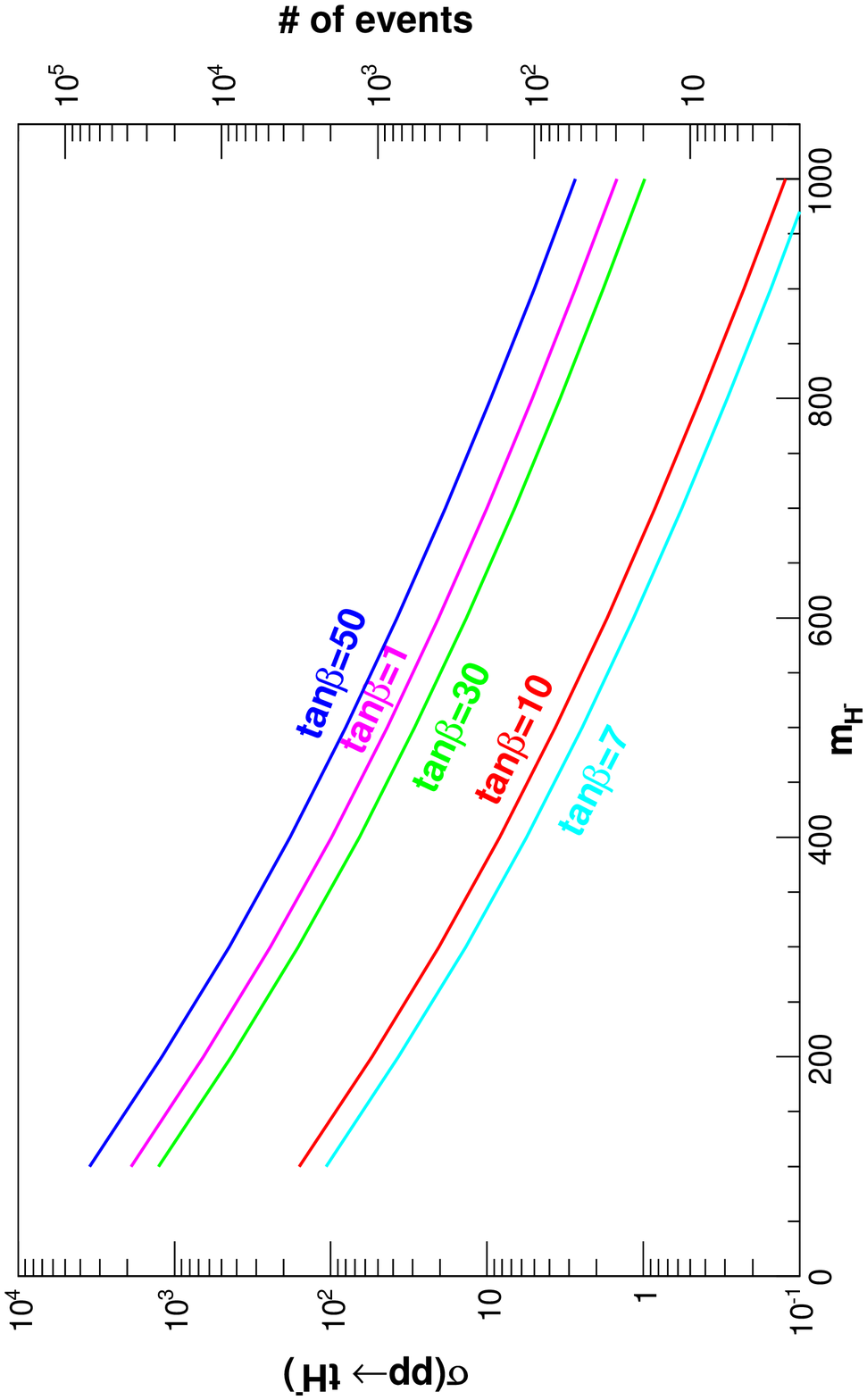} 
\end{tabular}}
\rotatebox{270}{
\begin{tabular}{c}
\includegraphics[width=0.30\textwidth]{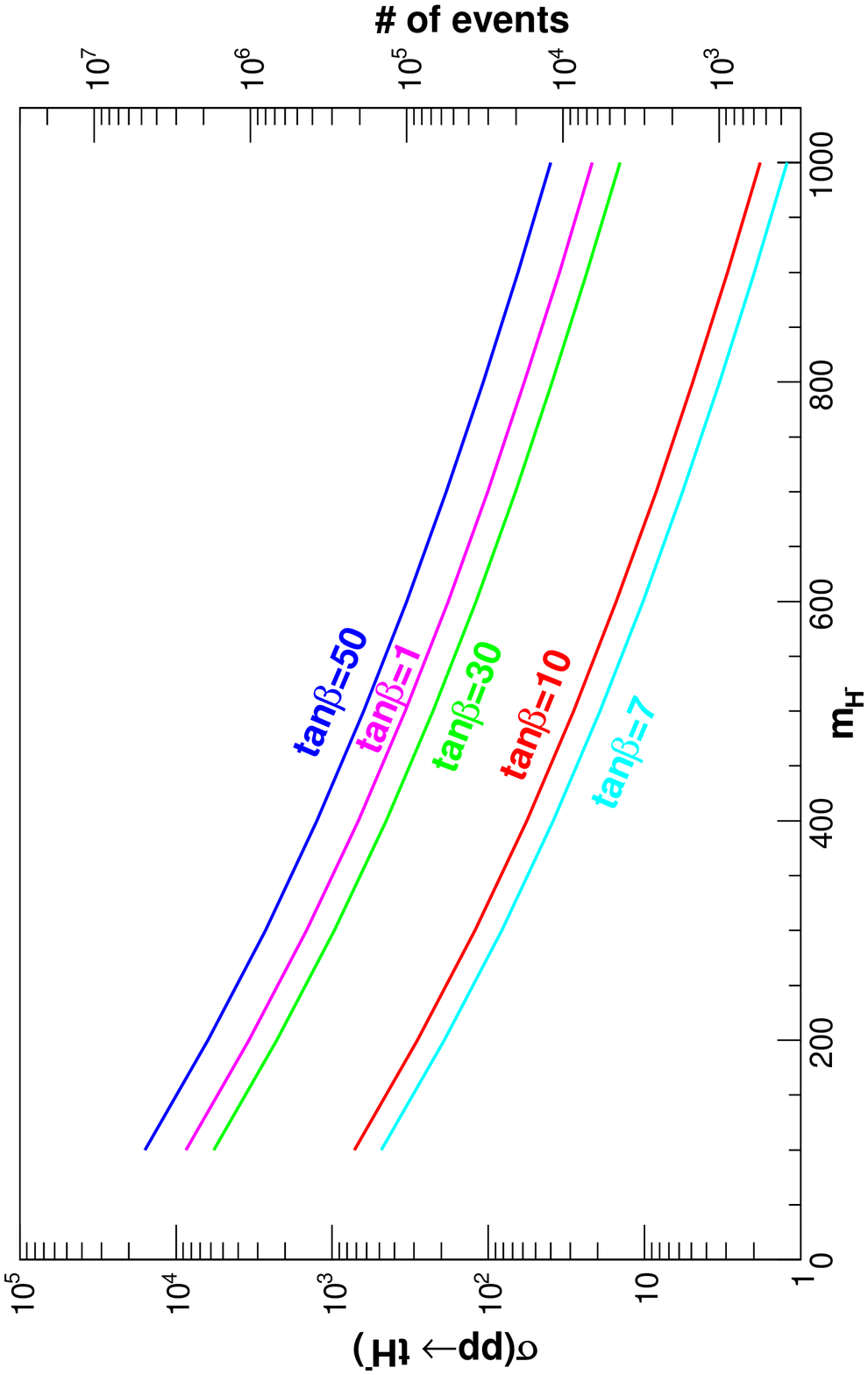}  
\end{tabular}}
\caption{The cross section and number of events for signal process $pp\to tH^{-}$ at centre of mass energy 8 TeV LHC with luminosity 20$fb^{-1}$ (left panel) and centre of mass energy 14 TeV LHC with luminosity 300$fb^{-1}$ (right panel).} 
\label{fig:cs-tH-} 
\par\end{centering}
\begin{centering}
\par\end{centering}
\centering{} 
\end{figure}
In Fig.~\ref{fig:cs-tH-}, the total cross sections and number of events for $pp\to tH^{-}$ are shown as a function of charged Higgs mass for
$\tan\beta=1, 7, 10$, 30, and 50 in 2HDM at the LHC with 8 TeV and 14 TeV. With assumed integrated luminosity to be 20 fb$^{-1}$ for 8 TeV and 300 fb$^{-1}$ for 14 TeV, it is shown that there are significant number of events for charged Higgs mass up to 1 TeV. Considering the decay modes of the top quark, we 
investigate the processes
\begin{equation}
pp\to tH^{-}\,\to\, t\bar{t}b\,\to\, bl^{+}\,+\, b\bar{b}jj\,+\,\met,
\label{tlhj}
\end{equation}
\begin{equation}
pp\to tH^{-}\,\to\, t\bar{t}b\,\to\, bjj\,+\, b\bar{b}l^{-}\,+\,\met.
\label{tjhl}\end{equation}
In process (\ref{tlhj}), the top quark produced associated with
$H^{-}$ decays semi-leptonically, and the anti-top quark from charged
Higgs decays hadronically, i.e., $t\to bl^{+}\nu_{l}$ and $\bar{t}\to\bar{b}jj$.
While in process (\ref{tjhl}), ${t}\to{b}jj$ and $\bar{t}\to\bar{b}l^{-}\bar{\nu}_{l}$.
The dominant SM background process is from $pp\to t\bar{t}j$.
To be more realistic, the simulation at the detector is performed
by smearing the energies of final state leptons and jets, according to the assumption
of the Gaussian resolution parameterization


For our signal process, one of the top quarks which decays hadronically can
be reconstructed from the final state three jets.
However, to reconstruct another top which decays leptonically, we have to utilize kinematical constraints to reconstruct its four-momentum because of the missing energy from neutrinos. The transverse momentum of neutrino can be obtained by momentum
conservation from the observed particles 
\begin{equation}
{\bf p}_{\nu T}=-({\bf p}_{lT}+\sum_{j=1}^{5}{\bf p}_{jT}),
\end{equation}
 while its longitudinal momentum can not be determined in this way
due to the unknown boost of the partonic  centre-of-mass system. Alternatively,
it can be solved with twofold ambiguity through the on-shell condition
of the W boson \begin{equation}
m_{W}^{2}=(p_{\nu}+p_{l})^{2}.\end{equation}
 Furthermore one can remove the ambiguity through the reconstruction
of the other top quark.  We thus evaluate the invariant
mass for each possibility\begin{equation}
M_{jl\nu}^{2}=(p_{l}+p_{\nu}+p_{j})^{2},\end{equation}
 where $j$ refers to the any one of the two left jets and pick up
the solution which is the closest to the top quark mass. With such a solution,
we can reconstruct the four-momentum of the neutrino and that of the left
top quark.

\begin{figure}
\centering \includegraphics[width=0.45\textwidth]{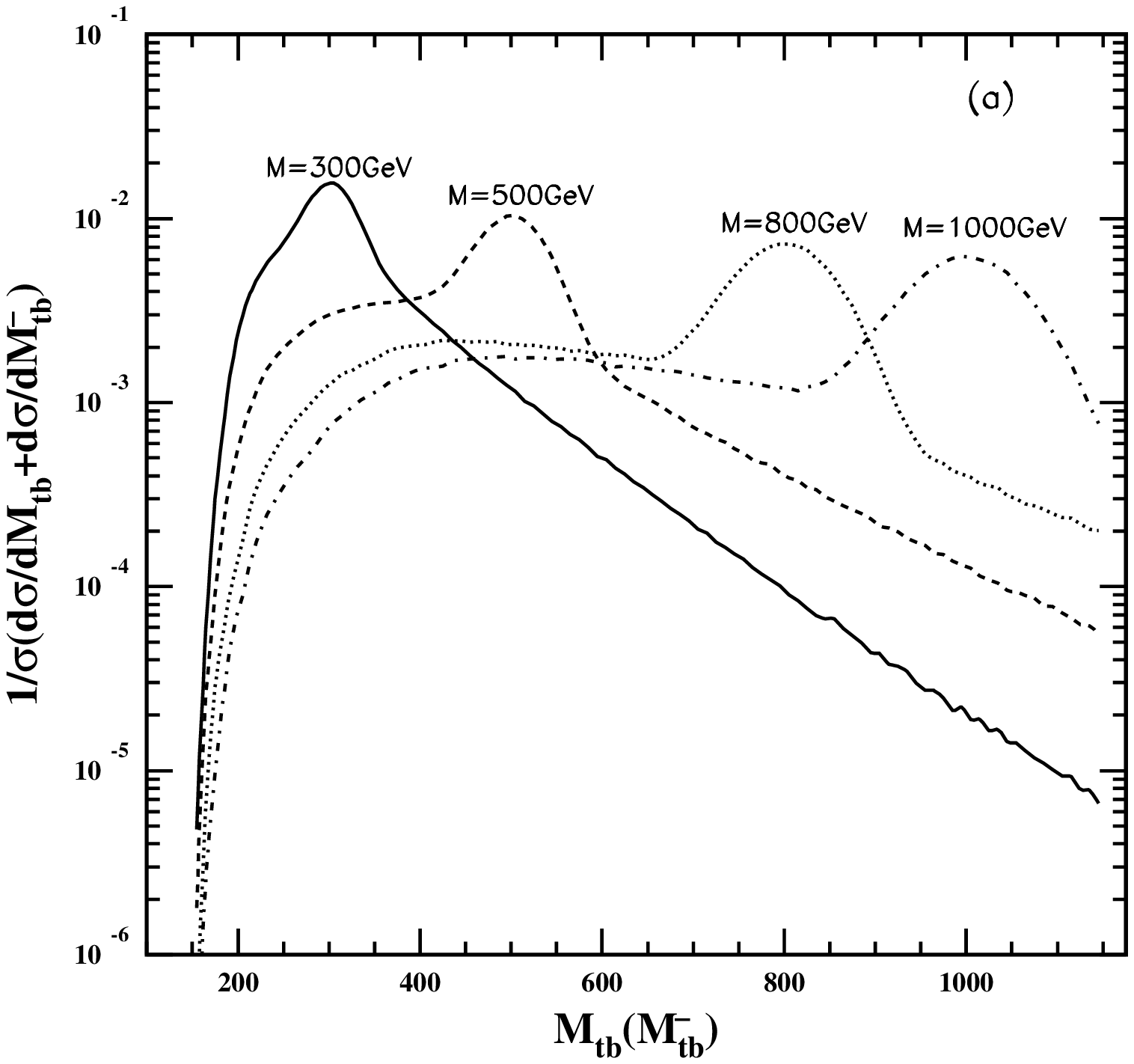} 
\includegraphics[width=0.45\textwidth]{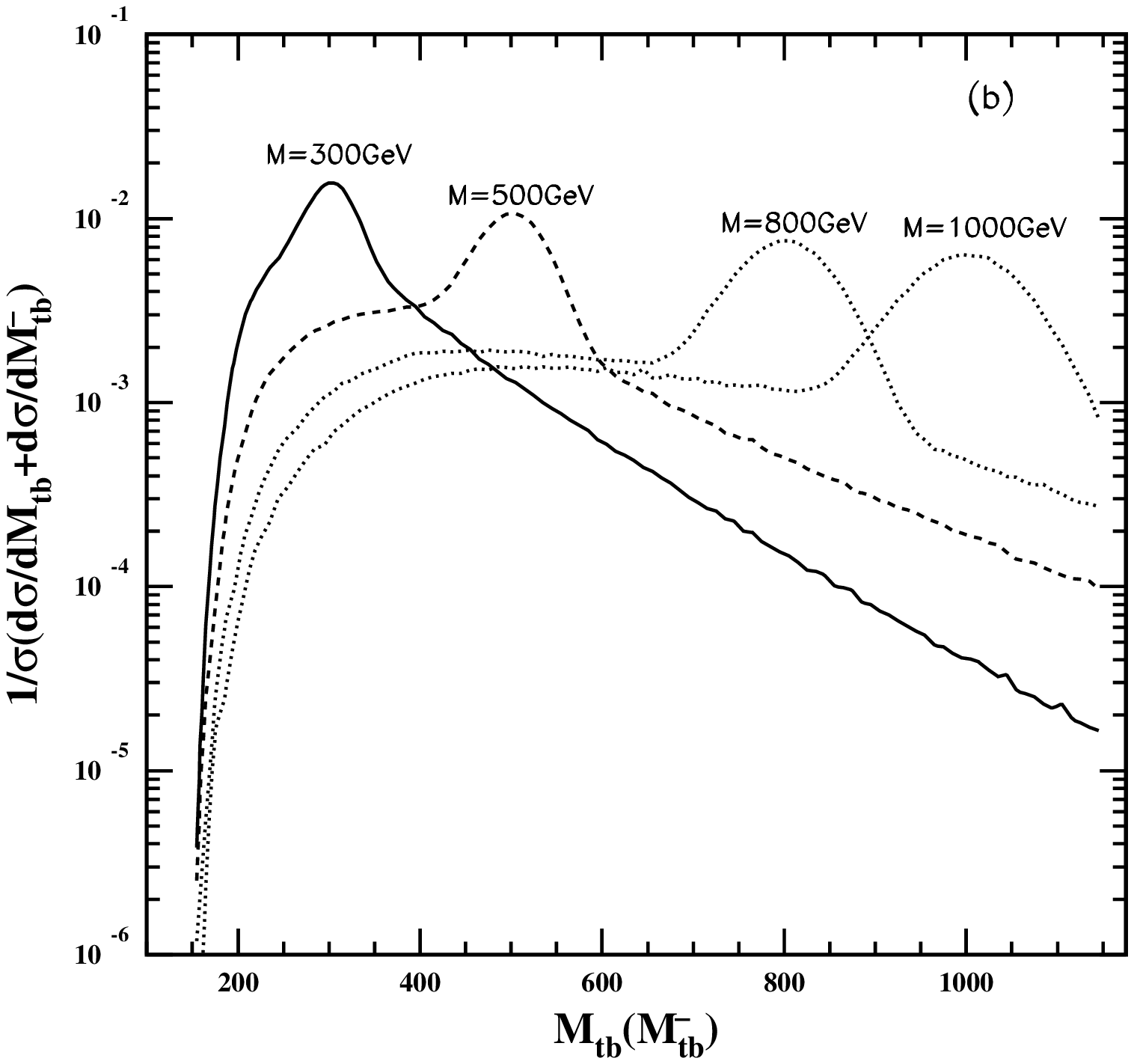}
\caption{The invariant mass distributions $1/\sigma(d\sigma/dM_{tb}+d\sigma/dM_{\bar{t}b})$.  $M_{tb} (M_{\bar{t}b})$ is the invariant mass between the reconstructed top (antitop) and the remaining jet 
for the process of $pp\to tH^{-}\to l^{+}+5jets+\met$ after cut I at LHC 
for (a) 8 TeV and (b) 14 TeV.}
\label{fig:mtb}
\end{figure}

Therefore, in the following numerical calculations, we apply the basic acceptance
cuts (cut I)
\begin{eqnarray}
 &  & p_{lT}>20~{\rm GeV},~~~~p_{jT}>20~{\rm GeV},~~~~\met>20~{\rm GeV},\nonumber \\
 &  & |\eta_{l}|<2.5,~~~~|\eta_{j}|<2.5,~~~~\Delta R_{jj(lj)}>0.4,\nonumber \\
 &  & |M_{j_{a}l\nu}-m_{t}|\leq30~{\rm GeV},~~~~|M_{j_{b}j_{c}j_{d}}-m_{t}|\leq30~{\rm GeV},\nonumber\\
&&|M_{j_{b}j_{c}}-m_{W}|<10~{\rm GeV}.
\label{cut1}\end{eqnarray}
After smearing and including the cut, the events were fully reconstructed ,
we can further reconstruct the invariant mass between the reconstructed top (antitop) 
and the remaining jet. We display the distributions  
$1/\sigma(d\sigma/dM_{tb}+d\sigma/dM_{\bar{t}b})$ in Fig.~\ref{fig:mtb},
where the resonance peaks are shown for different charged Higgs masses.
According to these resonance peaks we further employ a second cut (cut II)
%
\begin{itemize}
\item Cut II: $|M_{jj_{b}j_{c}j_{d}}-M|\leq10\%\, M$
or $|M_{jj_{a}l\nu}-M|\leq10\%\, M$.
\end{itemize}
Comparing our signal process with the dominant background process $pp\to t\bar{t}j$, after reconstructing the top and antitop quarks,
the remaining jet is a b-jet for our signal while it would probably be a light jet for the SM background process.
Therefore, we adopt the following cut ( cut III)
\begin{itemize}
\item Cut III: We demand the only remaining jet to be a b-jet.
\end{itemize}
The efficiency of $b$-tagging is assumed to be 60\% while the miss-tagging efficiency of a light jet as
a $b$ jet is taken to be transverse momentum dependent accordingly\cite{atlas0901}.

The cross sections and number of events for the signal and background processes (\ref{tlhj}) and (\ref{tjhl})
with different charged Higgs mass after each cut at
the LHC run of 8 TeV with integrated luminosity as 20 fb$^{-1}$ and 14 TeV  with integrated luminosity as 300 fb $^{-1}$ are respectively listed in Table.~\ref{tab:csa}.
The dominant SM background is
$pp\to t\bar{t}j\to l^\pm+5jets+\met$
process.  
At the LHC run of $\sqrt{s}=8$ TeV, we can see that it is difficult to detect such final states from $tH^{\pm}$ production when charged Higgs
mass is above 500 GeV. While with $300 fb^{-1}$ integrated luminosity at 14 TeV,
the significance can also be above three sigma for the charged Higgs mass $M\leq 1$ TeV. 
In the following, 
we will focus on the $tH^-$ production 
at  $\sqrt{s}=14$ TeV.
\begin{table}[h!]
 \tbl{The number of events of the signal process $pp\to tH^{-}\to l^{\pm}+5jets+\met$
and the background process of $pp\to t\bar{t}j\to l^{\pm}+5jets+\met$
at $\sqrt{s}=8$ TeV(left) and 14 TeV (right) before and after each cut.}{
\begin{tabular}{rl}
\begin{minipage}[t]{0.49\textwidth}\scalebox{0.79}{$
\begin{tabular}{|c||c|c|c|c|}
\hline
~Signal & \multicolumn{4}{c|}{$\sigma(pp\to tH^{-}\to t\bar{t}b\to l^{\pm}+5jets+\met)$ }\tabularnewline
\hline
$M$(TeV) & 0.3 & 0.5 & 0.8 & 1.0\tabularnewline
\hline
No cuts  &  903.6 &  172.4 &  20.4 &  6 \tabularnewline
\hline
Cut I   &   234.4 &  44 &  5 &  1.4  \tabularnewline
\hline
+Cut II &   191.8 &  34.6 &  4 &  1  \tabularnewline
\hline
+Cut III &    115.2 &  20.8 &  2.4 &  0.6\tabularnewline
\hline  \hline
Background & \multicolumn{4}{c|}{$\sigma(pp\to t\bar{t}j\to l^{\pm}+5jets+\met)$ } \tabularnewline
\hline
Cuts I+II+III &  205 &  77 &  19.4 &  9.2\tabularnewline
\hline  \hline
$S/B$         &  0.56 &  0.27 &  0.12 &  0.07    \tabularnewline
\hline
$S/\sqrt{S+B}$ &   6.43 &  2.10 &  0.51 &  0.19\tabularnewline
\hline
\end{tabular}$}
\end{minipage}
&
\begin{minipage}[t]{0.49\textwidth}\scalebox{0.79}{$
\begin{tabular}{|c||c|c|c|c|}
\hline
~Signal & \multicolumn{4}{c|}{$\sigma(pp\to tH^{-}\to t\bar{t}b\to l^{\pm}+5jets+\met)$ }
\tabularnewline
\hline
$M$(TeV) & 0.3 & 0.5 & 0.8 & 1.0 \tabularnewline
\hline
No cuts  &   78846  & 19788  & 3468  &  1284 \tabularnewline
\hline
Cut I    &    19737  & 4923  &  825  &  285\tabularnewline
\hline
+Cut II  &   16236  & 3900  &  660  &  231\tabularnewline
\hline
+Cut III  &   9741  &  2340  &  396  &  138\tabularnewline
\hline  \hline
Background & \multicolumn{4}{c|}{$\sigma(pp\to t\bar{t}j\to l^{\pm}+5jets+\met)$ } \tabularnewline
\hline
Cuts I+II+III &   13062  & 5691  &  1785  &  966 \tabularnewline
\hline  \hline
$S/B$        &     0.75 &  0.41 &  0.22 &   0.14\tabularnewline
\hline
$S/\sqrt{S+B}$ &   64.5 & 26.1 &  8.48 &   4.15\tabularnewline
\hline
\end{tabular}$}
\end{minipage}
\label{tab:csa}
\centering{}
\end{tabular}}
\end{table}

From Eqs. (\ref{eqst2}) and (\ref{eqst1}), we can see that the top
quark spin effect is related to the product of $g_{a}g_{b}$.
Using the same method as in Ref. 44 - 46, we find that this effect can be translated into the angular distributions of the charged lepton.
According to the (\ref{tlhj}) and (\ref{tjhl}), we obtain the angular distributions
\begin{eqnarray}
&&\frac{1}{\sigma}\,\frac{d\sigma}{d\cos\theta^{*}}\,=\,\frac{1}{2}\,[1\,+\, A_{FB}\cos\theta^{*}],~~~~\frac{1}{\sigma}\,\frac{d\sigma}{d\cos\bar{\theta}^{*}}\,=\,\frac{1}{2}\,[1\,+\,
\bar{A}_{FB}\cos\bar{\theta}^{*}],
\label{eqfbeq}\end{eqnarray}
 with
\begin{equation}
\cos\theta^{*}={{\bf\hat{p}}_{l^{+}}^{*}\cdot{\bf \hat{p}}_{t}^{*}},
~~~~~\cos\bar{\theta}^{*}={{\bf \hat{p}}_{l^{-}}^{*}\cdot{\bf \hat{p}}_{\bar{t}}^{*}},
\end{equation}
where ${\bf \hat{p}}_{l^{+}}^{*}$ is the 3-momentum unit vector of charged lepton in
the top quark rest frame, and ${\bf \hat{p}}_{l^{-}}^{*}$
is the 3-momentum unit vector of charged lepton in the antitop quark rest frame.
Without smearing effect and any acceptance cuts,
$A_{FB}$($\bar{A}_{FB}$) can be related to 
the top quark spin observables ${\cal{O}}_t$(${\cal{O}}_{\bar{t}}$), i.e.,
\begin{eqnarray}
A_{FB}=\kappa_c <{\cal O}_t>=2\kappa_c<\bf {{S_{t}}}\cdot  {\bf \hat p}_t^*>,\bar{A}_{FB}=\kappa_c <{\cal O}_{\bar{t}}>=2\kappa_c<\bf {{S}_{\bar t}}\cdot {\bf \hat p}_{\bar{t}}^*>.
\end{eqnarray}
Thus, for the charged lepton from semileptonic top quark decay, 
$A_{FB}=<{\cal O}_t>$ and $\bar{A}_{FB}= <{\cal O}_{\bar{t}}>$ before smearing effect and cuts.

According to Eq.(\ref{eqfbeq}), $A_{FB}$ and $\bar{A}_{FB}$ can also be determined as follows
\begin{eqnarray}
A_{FB}=\frac{\sigma(\cos\theta^{*}>0)-\sigma(\cos\theta^{*}<0)}
{\sigma(\cos\theta^{*}>0)+\sigma(\cos\theta^{*}<0)},\bar{A}_{FB}=\frac{\sigma(\cos\bar{\theta}^{*}>0)-\sigma(\cos\bar{\theta}^{*}<0)}
{\sigma(\cos\bar{\theta^{*}}>0)+\sigma(\cos\bar{\theta}^{*}<0)}.
\end{eqnarray}
These observables are useful to discriminate $Htb$ interactions from different models.  
In the framework of Type II 2HDM, the form of the Yukawa coupling is determined by
only one free parameter $\tan \beta$.  Combining the results of the $tH^-$ production cross section 
together with the forward-backward asymmetry  $A_{FB}$  and $\bar{A}_{FB}$, one can subtract the information of
$\tan\beta$ related to the $Htb$ coupling.
 Furthremore, we  extend our discussions in a 
general model where the Yukawa coupling of the charged Higgs bosons to fermions is
a free parameter so one can regard 
the scalar and pseudo-scalar parts of the Yukawa coupling as
completely independent and free parameters.
In the following, as an example,  we choose  $\tan\beta=30$ and
investigate the charged lepton angular distributions for three different
combinations of $g_{a} g_{b}$:

\begin{itemize}
\item $g_ag_b>0$,   e.g.,\\
$g_{a}=\pm g({\rm cot}\beta m_{t}+{\rm tan}\beta m_{b})/(2\sqrt{2}m_{{\rm W}})$, \\
$g_{b}=\pm g(\cot\beta m_{t}-\tan\beta m_{b})/(2\sqrt{2}m_{{\rm W}})$.
\item $g_ag_b=0$,  e.g.,\\ 
$g_{a}=0$, $g_{b}=g(\cot\beta m_{t}-\tan\beta m_{b})/(2\sqrt{2}m_{{\rm W}})$ or\\
$g_{a}=g({\rm cot}\beta m_{t}+{\rm tan}\beta m_{b})/(2\sqrt{2}m_{{\rm W}})$, $g_b=0$.
\item $g_ag_b<0$, e.g.,\\
$g_{a}=\pm g({\rm cot}\beta m_{t}+{\rm tan}\beta m_{b})/(2\sqrt{2}m_{{\rm W}})$,\\
$g_{b}=\mp g(\cot\beta m_{t}-\tan\beta m_{b})/(2\sqrt{2}m_{{\rm W}})$.
\end{itemize}
The predictions for $A_{FB}$ and $\bar{A}_{FB}$ corresponding to $g_ag_b$ are listed in Table \ref{tab:asy}.
Because the cross section contribution from the $s$-channel(Fig.\ref{fig:com1}(a))
decreases as the charged Higgs mass increases, without kinematical cuts,
the $\cos\theta^*$ distribution and
the $A_{FB}$ which are related to $tH^-$ production
 also depend on $M$; while the $\cos\bar{\theta}^*$ distribution and
the  $\bar{A}_{FB}$ which are related to the charged Higgs decay becomes $M$ independent.
More specific results can be found in Ref~47. From our study, after the acceptance cuts,
the angular distribution with respect to $\cos\theta^*$($\cos\bar{\theta}^*$) and $A_{FB}$($\bar{A}_{FB}$) is
more helpful to investigate the $Htb$ interaction for light(heavy) charged Higgs production associated with
top quark at the LHC.



\begin{table}[h!]
\tbl{The forward-backward asymmetry $A_{FB}$($\bar{A}_{FB}$)
for $pp\to tH^{-}\to l^{+}(l^-)+5jets+\met$ at LHC $\sqrt{s}=14$ TeV before
and after all cuts.}{
\scalebox{0.9}{$
\begin{tabular}{|c||c|c|c|c||c|c|c|c|}
\hline
 & \multicolumn{4}{c||}{$A_{FB}$}
 & \multicolumn{4}{c|} {$\bar{A}_{FB}$}\tabularnewline
\hline \hline
 \multicolumn{9}{|c|}{without cuts and smearing effect} \tabularnewline
\hline \hline
$M$(TeV) & 0.3 &0.5 & 0.8 & 1.0 & 0.3 & 0.5 & 0.8 & 1.0 \tabularnewline
\hline
$(g_ag_b)>0$  & 0.124  & 0.075  & 0.023  & -0.003  & -0.173  & -0.173 & -0.172  & -0.173\tabularnewline
\hline
$(g_ag_b)=0$  & 0.0 & 0.0  & 0.0  & 0.0  & 0.0  & 0.0  & 0.0  & 0.0   \tabularnewline
\hline
$(g_ag_b)<0$  & -0.125  & -0.076 & -0.024 & 0.001  & 0.172  & 0.173 & 0.174  & 0.173\tabularnewline
\hline  \hline
 \multicolumn{9}{|c|}{with cuts and smearing effect} \tabularnewline
\hline \hline
$(g_ag_b)>0$   & 0.002 & -0.015  & -0.031  & -0.041  & -0.014  & -0.050 & -0.054  & -0.061\tabularnewline
\hline
$(g_ag_b)=0$   & -0.028  & -0.030 & -0.033  & -0.041  & -0.022  & 0.007 & 0.006 & -0.006\tabularnewline
\hline
$(g_ag_b)<0$   & -0.056  & -0.048 & -0.037 & -0.040  & -0.033  & 0.081& 0.077  & 0.065\tabularnewline
\hline
\end{tabular}$}
\label{tab:asy}}
\end{table}

\section{Summary}\label{summary}

The observation of the charged Higgs is very important as an unambiguous signal for the existence
of new physics beyond SM. Therefore it is necessary to study the related
phenomena both theoretically and experimentally. In this letter, we study the determination of charged Higgs coupling of
the $tH^{-}$ associated production via $pp\to tH^{-}\to t\bar{t}b\to l^\pm +bb\bar{b}jj+\met$
process at LHC. 
We find that with 300 $fb^{-1}$ integral luminosity at
 $\sqrt{s}=14$ TeV, the signal
can be distinguished from the backgrounds for the charged
Higgs mass up to  1 TeV or  even larger.
If the $tH^-$ production
is observed at the LHC, one of the following questions is to identify
the $Htb$ interaction. For this purpose, we study  the angular
distributions of the charged leptons and the corresponding forward-backward
asymmetry induced by top quark spin. It is found that these distributions and
observables are sensitive to the product of $g_{a}g_{b}$, so that they can be
used to identify the $Htb$ interaction.
The $Htb$ interaction can be discriminated
with the help of the charged
lepton angular distribution and the forward-backward asymmetry
in the charged Higgs and top quark associated production at the LHC. 
Similar methods have also been proposed using such observables related to the top quark ~\cite{Cao:2013ud,Rindani:2013mqa}.
These analyses are helpful to distinguish the $Htb$ interaction in the 
2HDM or other new physics models with the presence of the charged Higgs.

\section*{Acknowledgments}

We would like to thank Profs. S.Y. Li, S.S. Bao and X.G. He for their helpful discussions and comments.
This work is supported in part by the National Science
Foundation of China (NSFC) under grant Nos. 11325525, 11275114, 11175251 and 11205023 and NSC.


\end{document}